\documentclass[namedreferences]{solarphysics}
\usepackage[hyperref,optionalrh,solaromanenum]{spr-sola-addons} 

\usepackage[utf8]{inputenc}
\usepackage[noenc]{tipa}
\usepackage{tipx}
\usepackage{pifont}
\usepackage{eurosym}
\usepackage{wasysym}
\usepackage{amssymb,amsfonts,textcomp}
\usepackage[T3,T1]{fontenc}
\usepackage[english]{babel}
\usepackage{color}
\usepackage{array}
\usepackage{hhline}
\hypersetup{colorlinks=true, linkcolor=blue, citecolor=blue, filecolor=blue, urlcolor=blue}
\usepackage{graphicx}
\usepackage{enumerate}
\usepackage{subfig}

\usepackage{caption}
\usepackage{lineno}

\usepackage{soul}

\usepackage{CJKutf8}
\usepackage{ucs}

\newcommand{\Chinese}[3]{%
    \csname CJK*\endcsname{UTF8}{bsmi}%
    \CJKchar{#1}{#2}{#3}%
    \csname endCJK*\endcsname
}
\newcommand{\Chi}[1]{%
    \csname CJK*\endcsname{UTF8}{bsmi}%
    #1%
    \csname endCJK*\endcsname
}

\begin{document}
\begin{article}

\defcitealias{2015AN....336..225N}{NN15}
\defcitealias{doi:10.1002/asna.201512193}{C15}
\defcitealias{STEPHENSON20151537}{S15}
\defcitealias{2013A&A...552L...3U}{U13}


\begin{opening}
\title{Do the Chinese Astronomical Records Dated AD 776 January 12/13 Describe an Auroral Display or a Lunar Halo? A Critical Re-examination.} 

\author[addressref={Durham},corref,email={f.r.stephenson@durham.ac.uk}]{\inits{F.R.}\fnm{F. Richard}~\lnm{Stephenson}}
\address[id=Durham]{Department of Physics, University of Durham, Durham, DH1 3LE, UK}

\author[addressref={STFC,Warwick},corref,email={david.willis@stfc.ac.uk}]{\inits{D.M.}\fnm{David M.}~\lnm{Willis}\orcid{0000-0001-9756-8601}}
\address[id=STFC]{UK Solar System Data Centre, Space Physics and Operations
  Division, RAL Space, Science and Technology Facilities Council, Rutherford
  Appleton Laboratory, Harwell Oxford, Didcot, Oxfordshire OX11 0QX, UK}
\address[id=Warwick]{Centre for Fusion, Space and Astrophysics, Department of Physics, University of Warwick, Coventry CV4 7AL, UK}

\author[addressref={STFC,Osaka}]{\inits{H.}\fnm{Hisashi}~\lnm{Hayakawa}\orcid{0000-0001-5370-3365}}
\address[id=Osaka]{Graduate School of Letters, Osaka University, Toyonaka, Osaka Pref. 5600043, Japan}

\author[addressref={RISH,USSS}]{\inits{Y.}\fnm{Yusuke}~\lnm{Ebihara}\orcid{0000-0002-2293-1557}}
\address[id=RISH]{Research Institute for Sustainable Humanosphere, Kyoto University, Gokasho, Uji, 6110011, Japan}
\address[id=USSS]{Unit of Synergetic Studies for Space, Kyoto University, Kitashirakawa-oiwake-cho, Sakyo-ku, Kyoto, 6068306, Japan}

\author[addressref={Reading}]{\inits{C.J.}\fnm{Christopher J.}~\lnm{Scott}\orcid{0000-0001-6411-5649}}
\address[id=Reading]{Department of Meteorology, University of Reading, Reading, RG6 6BB, UK}

\author[addressref={Zoo}]{\inits{J.}\fnm{Julia}~\lnm{Wilkinson}\orcid{0000-0001-8505-4494}}
\address[id=Zoo]{Zooniverse, c/o Astrophysics Department, University of Oxford, Oxford, OX1 3RH, UK}

\author[addressref={STFC}]{\inits{M.N.}\fnm{Matthew N.}~\lnm{Wild}\orcid{0000-0002-4028-2300}}

\runningauthor{F.R. Stephenson \textit{et al.}}
\runningtitle{Chinese Astronomical Records Dated  AD 776}

\begin{abstract}
The enhancement of carbon--14 in tree rings around AD 774/775 has generated wide interest in solar activity at that time. The historical auroral records have been examined critically. Of particular interest was the ``white vapour'' observed in China on AD 776 January 12/13. Both Usoskin \textit{et al.} (2013, \textit{Astron. Astrophys.} 55, L3; \citetalias{2013A&A...552L...3U}) and Stephenson (2015, \textit{Adv. Sp. Res.} 55, 1537; \citetalias{STEPHENSON20151537}) interpreted this record as an auroral display. Subsequently, Neuhäuser and Neuhäuser (2015, \textit{Astron. Nachr.} 336, 225; \citetalias{2015AN....336..225N}) proposed five “criteria” for the likeliness of aurorae and on this basis rejected an auroral interpretation. Instead, they interpreted it as a lunar halo, and suggested there were no auroral records as a proxy of solar activity in the interval AD 774--785. We consider if their ``lunar halo hypothesis'' and their auroral criteria could be of use in future researches on historical auroral candidates. We first show a counter-example for the lunar halo hypothesis from a parallel record on 1882 November 17, which was seen as a whitish colour, in the southerly direction, and near the Moon. We then consider \citetalias{2015AN....336..225N}’s criteria on colour, direction, and sky brightness and investigate other counter-examples from early-modern auroral observations. We also consider the extension of the white vapour in AD 776 according to the distribution of Chinese asterisms, and show that its large extension was inconsistent with the lunar halo hypothesis. Conversely, the streaks of white vapour penetrating the eight Chinese asterisms can be reproduced if we consider auroral-ray structures at altitudes between 97 km and 170 km, along geomagnetic field lines between the \textit{L}--shells \textit{L}=1.55 and 1.64. Our investigations show that we should consider candidate auroral records in historical documents not on the basis of the newly suggested \textit{a priori} criteria by \citetalias{2015AN....336..225N} but on all the available observational evidence.

\end{abstract}

\end{opening}

\section{Introduction}
\label{sec:intro}

Considerable interest has been generated by studies of the carbon--14 (\textsuperscript{14}C) record in annual tree rings by \citet{2012Natur.486..240M} and \citet{2013NatCo...4E1748M}. These authors noted sharp rises by around 1 percent in the \textsuperscript{14}C level in AD 774/5 and again in AD 993/4. The former event was particularly marked. The results obtained by \citet{2012Natur.486..240M} have been confirmed by independent measurements on tree rings from a variety of areas across the globe \citep[\textit{e.g.},][]{2013A&A...552L...3U,doi:10.1002/2014GL059874,2015E&PSL.411..290G,Buntgen2018,Uusitalo2018}. Further confirmation comes from \textsuperscript{14}C measurements of the corals of the South China Sea \citep{2014NatSR...4E3728L} and the \textsuperscript{10}Be concentration in the Antarctic Dome Fuji ice core \citep{2014AGUFMSH43A4185M,2015NatCo...6E8611M,2017JASTP.152...50S}. 

On the assumption that the observed \textsuperscript{14}C increases in AD 774/5 and AD 993/4 were of cosmic origin, a number of authors have searched for supporting evidence among historical astronomical records. Proposed mechanisms include: $\gamma$-rays from nearby supernova outbursts \citep{2012Natur.486..240M}; a comet entering the Earth’s atmosphere \citep{2014NatSR...4E3728L}; and solar flare events \citep{2012Natur.486..240M,2013A&A...552L...3U,2017SoPh..292...12H}. \citet{2015NatCo...6E8611M} examined not only \textsuperscript{14}C in tree rings but also \textsuperscript{10}Be and \textsuperscript{36}Cl in ice cores and detected peaks near AD 774/5 and AD 993/4 in all of these proxies, and concluded that these events are most likely to have been caused not by $\gamma$-ray bursts but by solar--proton events. 

In a search for astronomical evidence supporting the measured \textsuperscript{14}C increases in tree rings in about AD 774/5, various historical documents have been examined. \citet{2013A&A...552L...3U} surveyed celestial records in the historical documents around AD 774/5 and suggested enhanced solar activity in the mid-770s. \citet[hereafter S15]{STEPHENSON20151537} critically reviewed the available celestial records in contemporary historical documents. These documents included reports of supernovae, comets, sunspots, and aurorae. While not doubting the reliability of the impressive \textsuperscript{14}C results around AD 774/5 and AD 993/4, based on tree-ring measurements and other radioisotopes \citep[\textit{e.g.},][]{2012Natur.486..240M,2013NatCo...4E1748M,2015NatCo...6E8611M,Buntgen2018,Uusitalo2018}, \citetalias{STEPHENSON20151537} argued that the supporting astronomical evidence appeared to be far from conclusive. In particular, he stressed that one or two reliable sightings of the aurora over a period of a few years, even when fairly close in time to a significant increase in the \textsuperscript{14}C level, do not necessarily provide strong supporting evidence. For example, the approximate coincidence in date between the \textsuperscript{14}C events and auroral sightings could be largely coincidental. Importantly, however, \citetalias{STEPHENSON20151537} noted that the surviving historical records are far from complete; down the centuries several key celestial observations may have gone missing.

In the part of his review that considered possible auroral observations within several years of the \textsuperscript{14}C event occurring in AD 774/5, \citetalias{STEPHENSON20151537} concluded that there was only a single definite record of an auroral display in a dark sky, namely that recorded in China on a revised date corresponding to AD 776 January 12, and a plausible record from the Anglo-Saxon Chronicle \cite[see][]{Hayakawa2019}. Subsequently, \citet[hereafter NN15]{2015AN....336..225N} suggested five “criteria of likeliness” for auroral records for the auroral surveys around AD 774/5 and \citet[hereafter C15]{doi:10.1002/asna.201512193} criticised the interpretation of the Chinese record dated 776 January 12 as being a description of an auroral display, arguing instead that the record describes a lunar halo, based on these new criteria of likeliness.

The criticisms propounded by \citetalias{doi:10.1002/asna.201512193} of an interpretation in terms of a definite auroral display advanced by \citetalias{STEPHENSON20151537} may be summarised succinctly in terms of the following questions and assertions: i) Why was the date AD 776 January 12 not listed and classified as an aurora in the catalogue compiled by \citet{1995caof.book.....Y}?; ii) Why was the lunar halo hypothesis not considered?; iii) Why was no consideration given to the fact that the white \textit{qi} phenomenon (\textit{i.e.}, the ``white vapour'' described in the parallel records presented in \citet{1995caof.book.....Y} and \citet{2000eaaa.book.....X}) appeared in the eastern to southern direction, not in the north direction?; iv) The Chinese record is not consistent with an auroral display, partly because it would be too bright close to the Moon to observe the aurora; and v) The Chinese record could not describe an auroral display because the phenomenon was seen above the Moon.

It is important to note that these criticisms are to some extent ignoring the real context of \citetalias{STEPHENSON20151537} and end up reinforcing the main conclusion of \citetalias{STEPHENSON20151537}. In particular, \citetalias{STEPHENSON20151537} had argued conservatively that the available historical records are far from being entirely conclusive in supporting a solar origin of the spike in the \textsuperscript{14}C level. \citetalias{2015AN....336..225N} (without citing \citetalias{STEPHENSON20151537}) and \citetalias{doi:10.1002/asna.201512193} (citing \citetalias{STEPHENSON20151537}) criticised the few auroral candidates discussed in \citetalias{STEPHENSON20151537}, which would at best support the conservative discussions on contemporary historical documents by \citetalias{STEPHENSON20151537}. Therefore, removing the few definite or plausible auroral candidates would just reinforce the main conclusion in \citetalias{STEPHENSON20151537}. Nevertheless, it is still worth considering if the new criteria of likeliness and associated criticisms could be of any use in future searches for historical auroral candidates.

Therefore, the purpose of this paper is not to consider further the celestial observations recorded in East Asian and European history in the context of the measured increases in \textsuperscript{14}C around AD 774/5. Instead, the present paper focuses on the validity of the criticisms made by \citetalias{doi:10.1002/asna.201512193} of an auroral interpretation of the Chinese record dated 776 January 12 and, on the basis of \citetalias{doi:10.1002/asna.201512193}’s criticisms, the criteria of likeliness proposed by \citetalias{2015AN....336..225N}, which are listed explicitly in Section~\ref{sec:crcar}. In particular, it is argued that the description of the phenomenon observed on the night of AD 776 January 12/13, as recorded in the Chinese histories, is not consistent with the lunar halo hypothesis propounded by \citetalias{doi:10.1002/asna.201512193}. Moreover, attention is drawn to the potential implications for a substantial body of literature in the field of solar--terrestrial physics (STP), if all of the criticisms made by these authors were valid. While it is entirely legitimate and proper to challenge current ideas and conclusions, the concomitant consequences of the rejection of current ideas and conclusions must also be fully assessed.

\section{The Chinese Historical Texts: AD 776 January 12}
\label{sec:cht}

First, it is important to reproduce the original Chinese records that describe the phenomenon observed on the night of AD 776 January 12/13. Relevant historical information is recorded in the \textit{Jiu Tangshu} (v.36, p.~1328) and the \textit{Xin Tangshu} (v.32, p.~836), both of which are treatises on astronomy, while many auroral records in the \textit{Xin Tangshu} are found in treatises on the five elements. In the \textit{Jiu Tangshu} (“Old History of the Tang Dynasty”) the text is to be found in a general collection of astronomical records of various kinds of “portents (\textit{zaiyi})”. This older history was compiled between AD 940 and 945, soon after the end of the Tang Dynasty in AD 907. The \textit{Xin Tangshu} (“New History of the Tang Dynasty”) was compiled more systematically between AD 1043 and 1060. The older history (\textit{Jiu Tangshu}) often contains details that are omitted in the newer history (\textit{Xin Tangshu}), for example, a few eclipses \citep[\textit{e.g.},][p.~245]{1997heer.book.....S}. Whereas the astronomical treatise of the \textit{Jiu Tangshu} tends to list events purely in chronological order, regardless of their type, the section of “portents (\textit{zaiyi})” in the astronomical treatise of the \textit{Xin Tangshu} divides the entries into groups: for example, solar eclipses, solar omens, lunar omens, comets, \textit{etc.} However, the scientific significance of these divisions is at least debatable.

A full presentation and translation of the record in the \textit{Jiu Tangshu} (v.36, p.~1328) is as follows:
\begin{quotation}
    \Chi{大曆十年…十二月丙子夜，東方月上有白氣十餘道，如匹帛，貫五車、東井、輿鬼、觜、參、畢、柳、軒轅，三更後方散。}
    
    ``Dali reign period, 10th year, 12th lunar month, day \textit{bingzi} (12) [= AD 776 Jan 12]. At night, in the E direction above the Moon there were more than ten streaks (\textit{dao}) of white vapour (\textit{baiqi}). They were like unspun silk. They penetrated (\textit{guan}) [the star groups] \textit{Wuche}(in Auriga), \textit{Dongjing} (in Gemini), \textit{Yugui} (in Cancer), \textit{Zui} and \textit{Shen} (both in Orion), \textit{Bi} (in Taurus), \textit{Liu} (in Hydra), and \textit{Xuanyuan} (in Leo). After the third watch (\textit{i.e.}, after about 1:30 a.m), they disappeared.''
\end{quotation}

A full presentation and translation of the record in the \textit{Xin Tangshu} (v. 32, p.~836) is as follows:
\begin{quotation}
    \Chi{大曆十年…十二月丙子，月出東方，上有白氣十餘道，如匹練，貫五車及畢、觜觿、參、東井、輿鬼、柳、軒轅，中夜散去。}
    
    “Dali reign period, 10th year, 12th lunar month, day \textit{bingzi} (12) [= AD 776 Jan 12], the Moon appeared (=rose) in the eastern direction. Above it there were more than ten streaks (\textit{dao}) of white vapour (\textit{baiqi}). They were like unspun silk. They penetrated \textit{(guan}) [the star groups] \textit{Wuche} and \textit{Bi}, \textit{Zuixi}, \textit{Shen}, \textit{Dongjing}, \textit{Yugui}, \textit{Liu}, and \textit{Xuanyuan}. After midnight they disappeared.”
\end{quotation}

It is not absolutely certain from the historical records if the “more than ten streaks of white vapour” penetrated the eight Chinese star groups simultaneously soon after moonrise, or possibly sequentially during the course of the night up to about 1:30 a.m. In this context, it is perhaps important to note the difference in the order of the star groups in the two histories. As a further (minor) distinction between the two histories, the \textit{Jiu Tangshu} gives “\textit{Zui}”, whereas the \textit{Xin Tangshu} gives the full name “\textit{Zuixi}”.

However, it should be noted that the description of the white vapour is remarkably detailed, considering it was recorded in the 8th century. The observer has recorded white vapour penetrating more than 8 asterisms (Chinese constellations). The stars in some of these asterisms are quite faint in terms of their magnitude, as in \textit{Yugui} (${\geq}$ 3.9), \textit{Zuixi} (${\geq}$ 3.4), or \textit{Liu} (${\geq}$ 3.1) \citep[see][]{ref10140349,nai1989zhongguo}. This shows the observer had some specific skill and experience of astronomical observations.

The site at which these observations were made is considered to be the imperial observatory, due to the tradition of astronomical observations being made by official astronomers \citep[\textit{e.g.},][]{Keimatsu1970,pankenier_2013,Hayakawa2015,2017PASJ...69...65H,doi:10.1093/pasj/psx087}. It is recorded that the astronomical observatory (\textit{sitiantai}) was moved to \textit{Yongningfang} District in 758 (\textit{Jiu Tangshu}, v.36, p.~1335). According to the city plan of Chang’an during the \textit{Tang} Dynasty (\textit{e.g., Tang Lianjing Chengfangkao}, v.1, ff.1b-2a; v.3, ff.11b-13a), it is possible to identify the observatory location as 34°14$'$N, 108°56$'$E.

\section{Criteria for the Reliability of Candidate Auroral Records}
\label{sec:crcar}

In a discussion of solar activity around AD 775, based on contemporaneous observations of aurorae and radiocarbon, \citetalias{2015AN....336..225N} have proposed five criteria for the \textit{likeliness} of an event included in an auroral catalogue to be an aurora. These criteria were selected to distinguish likely aurora from other phenomena (\textit{e.g.}, halos, eclipses, meteors, rainbows, volcanic eruptions, comets, novae, weather phenomena, \textit{etc.} ). \citetalias{2015AN....336..225N} classify candidate auroral sightings as \textit{almost certain} ($N=5$), \textit{very probable} ($N=4$), \textit{probable} ($N=3$), \textit{very possible} ($N=2$), \textit{possible} ($N=1$) or \textit{potential} ($N=0$), where $N$ denotes the number of the following five criteria fulfilled:

\begin{enumerate}[i)]
    \item Colour, \textit{e.g.} reddish, fiery, blood(y), scarlet, green, blue, or the report of a \textit{dragon}/\textit{snake} or \textit{war armies} (both assumed to have some non-white colour), while wordings such as \textit{white}, \textit{brilliant}, or \textit{glow} are neither sufficient nor a contradiction (neutral);

    \item Aurora-typical motion (changes, pulses, or strong dynamics) including apparent motion indicated by the words \textit{fire}, \textit{fiery}, \textit{fight}, \textit{(war) armies}, \textit{dragon(s)};

    \item Northern direction (or intrinsically north by mentioning northern celestial regions) including NE and NW as well as from \textit{E to W} or \textit{W to E} (reports like, \textit{e.g., in the east} are neutral, but excluding purely southern sightings);

    \item Night-time observation (darkness), wordings like \textit{after sunset} or \textit{before sunrise} would not necessarily indicate aurorae, but also do not disprove the possibility of an aurora, so that they are considered neutral (when medieval authors write, \textit{e.g., after sunset}, they mean shortly after, if not even during sunset, but not during the deep dark night). If the event clearly happened during civil or nautical twilight, it cannot be concluded that it was an aurora; when the time of observation was given as Chinese night watches, this time has been converted to local time; and

    \item The event was repeated (even if weaker) in at least one of the next three nights.
\end{enumerate}

The five criteria introduced by \citetalias{2015AN....336..225N} are reproduced essentially verbatim in the preceding paragraph, in order to illustrate the differences compared to the earlier classification criteria used by \citet{1956JGR....61..297M} and \citet{Keimatsu1970}. However, although these more recent criteria may be adequate for ``normal'' aurorae, which cannot be classified as extreme events, they appear to be inadequate in the case of aurorae associated with extreme magnetic storms, such as the possible extreme magnetic storm around AD 775 -- according to the observed enhancements in cosmogenic isotopes \citep{2012Natur.486..240M,2013A&A...552L...3U,2015NatCo...6E8611M}. For example, if criteria i), ii), iv) and v) were all satisfied, then an aurora seen in the north (N) would achieve a score of 5, whereas an aurora seen only in the south (S) during an extreme event (perhaps as a result of extensive cloud cover in the N) would be excluded because \citetalias{2015AN....336..225N} are “excluding purely southern sightings”. This method of scoring seems slightly illogical and the associated shortcomings in the criteria defined by \citetalias{2015AN....336..225N} are considered further in the following sections.

It is speculated that \citetalias{2015AN....336..225N} restricted their “criteria of likeliness” to the reliability of observations of the \textit{aurora borealis} because they were considering reports of aurorae observed in East Asia and Europe. At least, \citetalias{2015AN....336..225N} did not specify the scope for their survey of records describing the \textit{aurora borealis}. In particular, \citetalias{2015AN....336..225N} did not explicitly restrict their criteria to the \textit{aurora borealis}, and appear to have overlooked the fact that the \textit{aurora australis} will be more likely to appear in the southward direction in the southern hemisphere \citep{1996QJRAS..37..733W,2009A&G....50e..20W}. 

It is appropriate to address at this point the first of the questions asked by \citetalias{doi:10.1002/asna.201512193}; namely: ``Why was the date AD 776 January 12 not listed and classified as an aurora in the catalogue compiled by \citet{1995caof.book.....Y}?'' Unfortunately, the Chinese record dated AD 776 January 12 was overlooked in the auroral catalogue compiled by \citet{1995caof.book.....Y}, at least partly because it was also overlooked in the catalogues compiled by \citet{dai1980historical} and the \citet{Beijing1988}.

\section{A Counter-example Refuting the Lunar Halo Hypothesis}
\label{sec:crlhh}

\begin{figure}
    \centering
    \includegraphics[width=15.055cm,height=8.645cm]{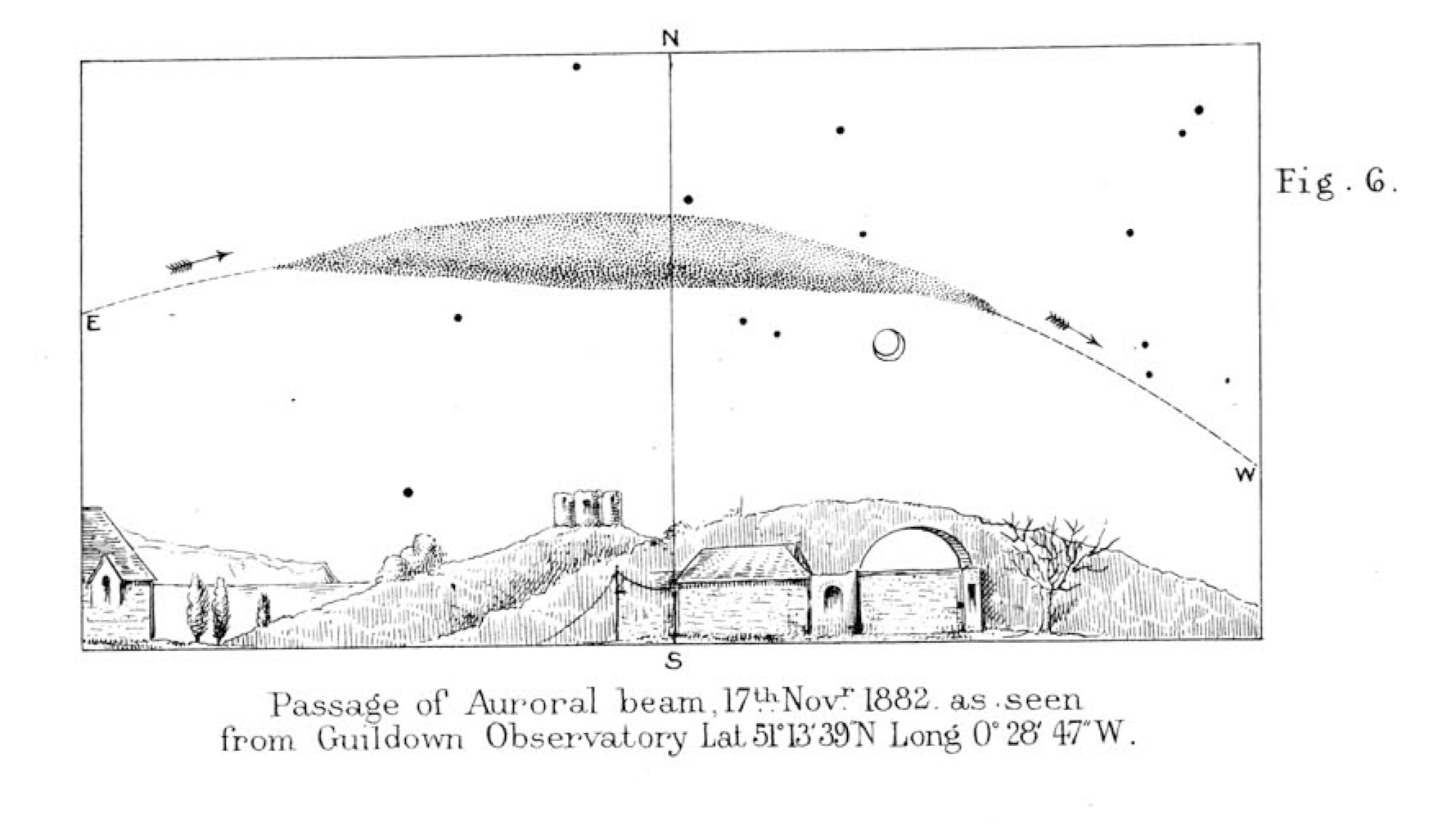}
    \caption{Passage of auroral beam on 1882 November 17 reproduced from \citet{Capron1883}. This auroral beam was seen in the “south”, was “pearly white” in colour, and was “in front of or near the moon” in position.}
    \label{fig:capron}
\end{figure}

The scientific reasons why \citetalias{2015AN....336..225N} and \citetalias{doi:10.1002/asna.201512193} relate the astronomical records in 776 not with an auroral display but with a lunar halo are: i) its direction was in the south (not in the north); ii) its location was near the Moon; and iii) its colour was whitish. Therefore, on the basis of the criteria of likeliness presented by \citetalias{2015AN....336..225N}, which are reproduced in Section~\ref{sec:crcar}, an auroral interpretation of the Chinese records dated 776 January 12/13 was rejected by \citetalias{doi:10.1002/asna.201512193}. However, an early-modern auroral observation casts serious doubt about the validity of the criteria of likeliness in assessing auroral observations recorded during strong space-weather events. On 1882 November 17, a great geomagnetic storm was recorded at Greenwich \citep{RGO1955} and its maximum Dst value was subsequently estimated to be $-386$nT, according to contemporary magnetic observations \citep{doi:10.1002/2017SW001795}. \citet{Capron1883}, a spectroscopist of the late 19th century, had observed this auroral event and reported an “auroral beam” in the southern sky near the Moon at Guildown Observatory (51°13$'$39$''$N, 0°28$'$47$''$W), Surrey, as reproduced in \autoref{fig:capron}. \citet[p.~319]{Capron1883} reported his observation as follows: ``About 6 P.M., while the aurora was fitfully blazing in the north, north-east, and north-western sky, in the east there rose from the horizon a long beam of detached bright light, which, apparently lengthening as it advanced, crossed rapidly the southern horizon in front of or near the moon, and then sank in the west, shortening in length as it did so. The light omitted from it was described by one observer as of a glowing pearly white''. \citet[p.~320]{Capron1883} confirmed that auroral beam was ``about 2 degrees above the Moon'' in terms of its angular separation.

Clearly, the main characteristics of the “auroral beam” observed on 1882 November 17 were: i) a direction in the south (not in the north), ii) a location near the Moon, and iii) a whitish colour. Other contemporary observations also described this aurora as being: ``quite close above the Moon''; ``aurora crossing the Moon''; aurora being ``just over the Moon, a green arch''; and a ``stationary arc 30 degrees to 40 degrees above horizon and above the Moon'' \citep[pp.~320--323]{Capron1883}. Therefore, if the discussions in \citetalias{2015AN....336..225N} and \citetalias{doi:10.1002/asna.201512193} were completely correct, this auroral beam should be interpreted not as an auroral phenomenon but as a lunar halo.

Fortunately, however, \citet{Capron1883} observed this auroral beam with his spectroscope and found the existence of the principal citron auroral line (W.L. 5569) and the absence of the “Fraunhofer dark lines”. The absence of the “Fraunhofer dark lines” indicates the observed phenomenon was not caused by reflected solar light, which rules out an explanation in terms of atmospheric optics such as a lunar halo. \citet{Capron1883} described the detail as follows: “In my own case (obs. 2) I was using the spectroscope (a large Browning direct-vision one) upon the aurora northwards, but, accidentally turning and seeing the beam, I at once applied the instrument to it. The spectrum (not to be found on the adjacent sky) was observed to consist of the well-known principal citron auroral line (W.L. 5569), and a faint greenish-white continuous spectrum extending from about D to F [the Fraunhofer spectral lines]. No other bright line than the principal one was visible; and the continuous spectrum showed no trace whatever of the Fraunhofer dark lines --- indicating an absence of solar reflected light.” The second observation was made by Mr. F. W. Cory, of Buckhurst Hill, Essex, who in a letter to 'Knowledge' (vol. iii. no. 62) says: --- “I think there can be no doubt in regard to the connexion between the torpedo-shaped body that was seen on November 17 at 6 h 5 m P.M. and the aurora, as the spectroscopic examination gave the same line for both; and this was situated between D and E in the spectrum, but nearer the former” \citep[p.~324]{Capron1883}. 

The report by \citet{Capron1883} provides clear evidence of the existence of an auroral beam on 1882 November 17 with: i) a whitish colour, ii) a location near the Moon, and iii) a southerly direction (not northerly). This evidence provides a single irrefutable counter-example of the claim in the papers by \citetalias{2015AN....336..225N} and \citetalias{doi:10.1002/asna.201512193} that the Chinese record of AD 776 January 12 cannot be interpreted as an auroral display because it was seen in the south, so consideration should be given instead to an interpretation in terms of a lunar halo. The only way of circumventing the criticism presented here would be to argue that the Chinese record of AD 776 January 12 was acquired at a time of moderate magnetic disturbance, not during a great geomagnetic storm, so that the five criteria of likeliness of a candidate record in an auroral catalogue being a true auroral event (\citetalias{2015AN....336..225N}) should be satisfied (see Section~\ref{sec:crcar}). Unfortunately, there is no direct evidence regarding the level of magnetic disturbance on AD 776 January 12, although the spatial extent of the more than ten streaks of white vapour suggests that a major magnetic storm may have been in progress (see Section~\ref{sec:jaicr}). Therefore, it is necessary to consider further examples of aurorae observed in non-northerly directions, aurorae seen near the Moon, and aurorae having a whitish colour, by examining other records of early-modern auroral observations.

\section{Aurorae Observed in Non-northerly Directions}
\label{sec:aonnd}

This section begins by addressing one of the questions that \citetalias{doi:10.1002/asna.201512193} posed on the paper by \citetalias{STEPHENSON20151537}, namely:  “Why was no consideration given to the fact that the white~\textit{qi}~phenomenon appeared in the eastern to southern direction, not in the north direction?” As noted in the previous section, the implication here appears to be that a true auroral display would be more likely to have been seen in a northerly direction (N, NE, or NW), rather than in a non-northerly direction (SE, S, SW), at least for observations made at middle to low latitudes. This conclusion follows from the five criteria of likeliness presented by \citetalias{2015AN....336..225N}, as discussed in Section~\ref{sec:crcar}. Of course, it is magnetic latitudes, not geographic latitudes, that are relevant in the case of aurorae and therefore it is important to consider the configuration and orientation of the geomagnetic field at the epoch an auroral display was observed. There is no overriding physical reason why aurorae could never be seen in non-northerly directions at Chang’an (34°14$'$N, 108°56$'$E).

In general, the auroral oval moves significantly equatorward at the times of great geomagnetic disturbance \citep{1998AnGeo..16..566Y,Cliver2004}. The lowest latitude at which the aurora was observed for certain extreme events has received much attention in the~scientific~literature \citep{Cliver2004,2006AdSpR..38..130G,2006AdSpR..38..136S,2008JASTP..70.1301S,2016PASJ...68...33H}. \citet{Cliver2004} place the extreme space–weather events on 1859 September 1/2 and 1872 February 4/5 at the very top of their list of the six most equatorward boundaries of auroral visibility from 1859 to 1958 in their Table 7.

The former space-weather event is the Carrington Event in 1859, arguably one of the greatest and most famous space$-$weather events in the last two centuries \citep{Kimball1960,doi:10.1029/2002JA009504,2006AdSpR..38..130G,2013JSWSC...3A..31C,2016PASJ...68...33H} and was associated with the white-light flare that occurred on the Sun on 1859 September 1 \citep{Carrington1859,Hodgson1859}. At their peak, the auroral displays were visible down to Honolulu (20.5° MLAT) on September 1/2. Taking the elevation angle of the auroral display at Honolulu to be 35° \citep[p.~88]{1860AmJS...30...79L}, we can estimate the equatorward boundary of the auroral display as 25.1° MLAT (28.5° ILAT for the magnetic footprint), assuming the altitude of the aurora to be 400 km \citep{1998JASTP..60..997S,2017SpWea..15.1373E}. This is consistent with the recent studies revealing that the auroral display extended up to the zenith at Mexico City (28.8° MLAT) \citep{doi:10.1029/2017SW001789,2018ApJ...862...15H}.

The latter space–weather event is the great magnetic storm of 1872 February 4/5, highlighted by \citet{2008JASTP..70.1301S} and \citet{2018ApJ...862...15H} as being comparable to, and perhaps even greater than, that of 1859 September 1/2 (the ‘Carrington Event’). In terms of its visibility, the auroral display on 1872 February 4/5 was essentially seen worldwide. The auroral observations extended to magnetic latitudes as low as 10°, in the Caribbean, Egypt, Southern Africa, the Indian Ocean, the Indian subcontinent, and China. In addition, \citet{2008JASTP..70.1301S} presented a report of an auroral observation from Bombay at 10.0° MLAT, citing \citet{1957Natur.179....7C}. Considering that this auroral display was seen almost in the zenith at Shanghai \citep[p.~8]{Donati1874}, the equatorward boundary of the auroral oval can be estimated as 20° MLAT (24.2° ILAT for the magnetic footprint) \citep[see,][]{2018ApJ...862...15H}.

\begin{figure}
    \centering
    \subfloat{%
        \label{fig:cambridge}%
        \includegraphics[width=0.45\textwidth]{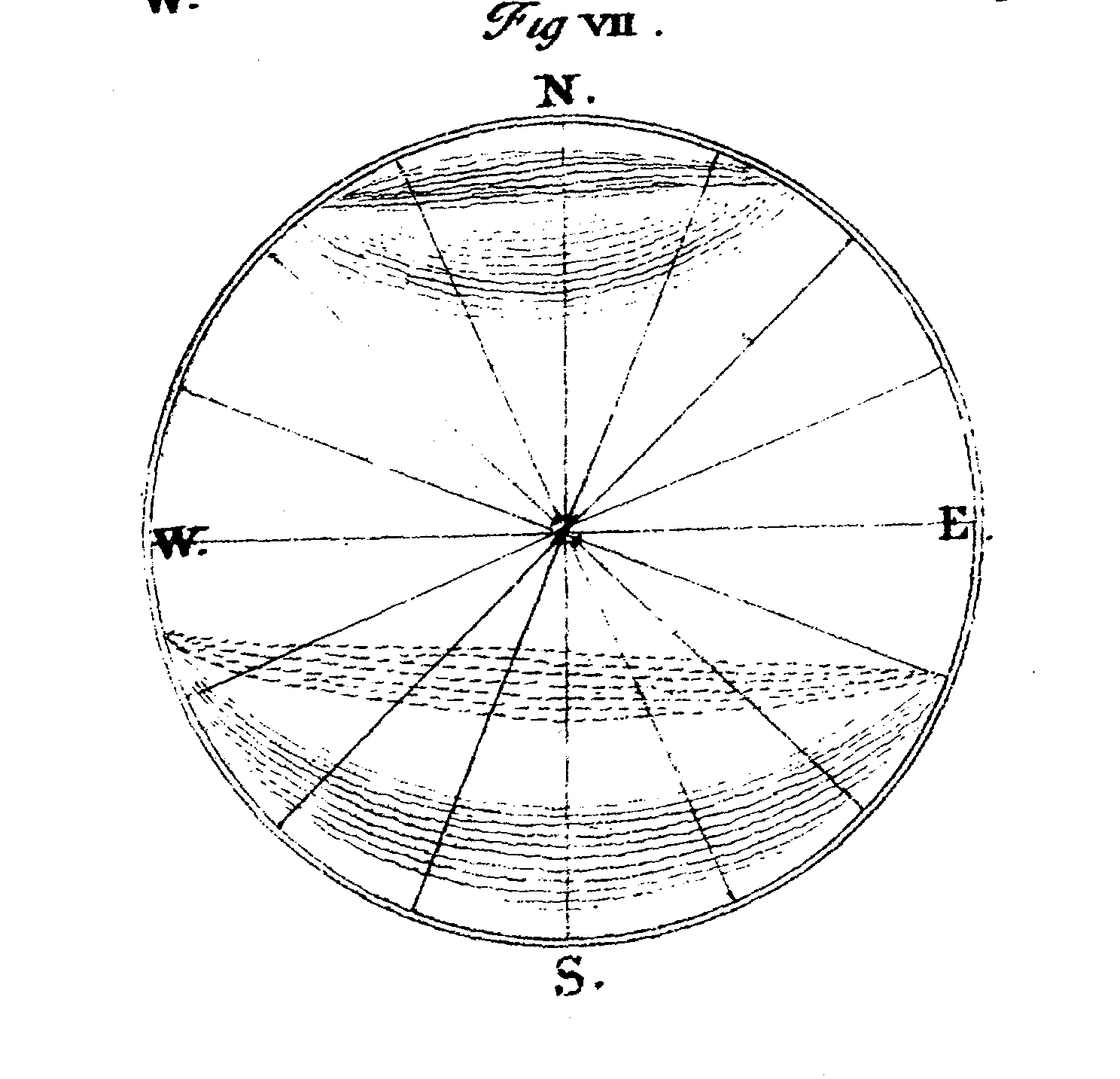}}
    \qquad
    \subfloat{%
        \label{fig:marseille}%
        \includegraphics[width=0.45\textwidth]{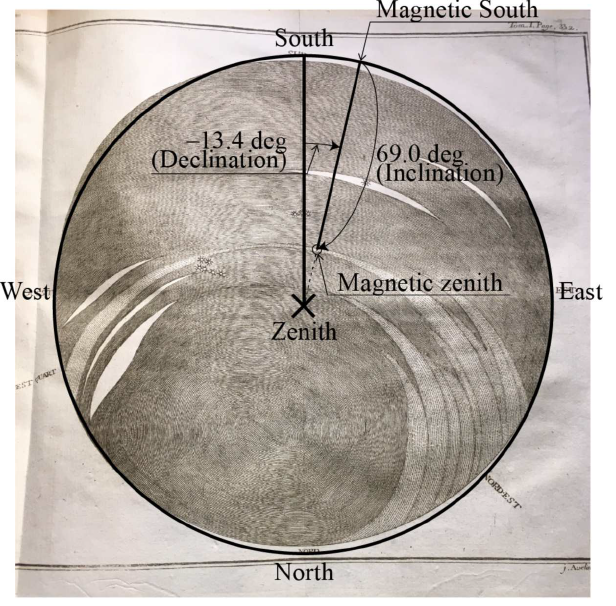}}
    \caption{(a, left) The auroral display observed at 22:25 LT at Cambridge on 22 October 1730 reproduced from \citet{ref2032076}; (b, right) The auroral display observed at Marseille on 15 February 1730, reproduced from \citet{Caumette1766}, as annotated by \citet{2018A&A...616A.177H}.}
    \label{fig:display}
\end{figure}

Even in the case of less extreme magnetic storms, some auroral displays extend beyond the zenith and are more splendid in a non-northerly direction. On 1730 October 22, a great auroral display covered the sky of the Northern Hemisphere. Isaac Greenwood, a contemporary professor of mathematics and natural philosophy at Cambridge in New-England (42°22$'$N, 71°07$'$W), reported the detailed appearance of auroral displays with their drawings \citep[pp.65--66]{ref2032076}. The auroral displays were notably splendid at 22:25 LT. He provides a figure showing “Observation XXV. 10\textsuperscript{h} 25$'$ Fig. 7, in which Z denotes the Zenith, and N. E. S. W. the Horizon”, as reproduced in Figure~\ref{fig:cambridge}. This figure indicates quite clearly that the auroral display in the southern sky was even more splendid than, and separate from, that in the northern sky. Therefore, it is concluded that the auroral displays can sometimes be more splendid in non-northerly directions of the sky, provided the associated magnetic storm is sufficiently strong.

Another case is the great magnetic storm on 1730 February 15, as reviewed by \citet{2018A&A...616A.177H}. \citet[pp.~4--5]{ref165260101} recorded that the illuminated region of the sky was inclined from west to south in Rome. On the same night, this auroral display was observed at Marseille as reproduced in Figure~\ref{fig:marseille}. In this figure, we also find the auroral display extended beyond the zenith and a blank space is found around the geographical true north. Note that the computed magnetic latitudes of Cambridge (42°22$'$N, 71°07$'$W), Rome (41°53$'$N, 12°29$'$E), and Marseille (43°18$'$N, 7°28$'$E) are 51.1° MLAT, 45.1° MLAT, and 47.6° MLAT, respectively, according to the geomagnetic field model GUFM1 \citep{Jackson2000}.

An auroral event on the night of AD 776 January 12/13 would have occurred more than 900 years before the great magnetic storms of the eighteenth and nineteenth centuries. Therefore, as already mentioned, it is imperative to consider the configuration and orientation of the main geomagnetic field this far back in time. It seems likely from the work of \citet{doi:10.1111/arcm.12152} that the geomagnetic north pole in 776 was located at 86°44$'$N, 79°08$'$E, and hence the MLAT of Chang’an (34°14$'$N, 108°56$'$E) can be estimated as 37.0° MLAT, according to the dipole axis determined by the global geomagnetic field model CALS3k.4b of \citet{Korte2011}. Indeed, Chang’an was at notably higher MLAT in 776 than in modern times \citep[24.5° MLAT in 2014; see,][]{2015EP&S...67...79T}.

From about AD 1000 to AD 1300, the geomagnetic north pole moved eastward and arrived at 88°14$'$N, 153°47$'$E in 1128 \citep{Korte2011,doi:10.1111/arcm.12152}. Indeed, Korean astronomers at Songdo (37°35$'$N, 126°18$'$E, 39.5° MLAT) recorded an auroral display on 1128 December 13 (\textit{Goryeosa}, v.53, f.54a), following the earliest known datable drawing of sunspots in the Chronicle of John of Worcester on 1128 December 8 \citep{1999A&G....40f..21S,2001AnGeo..19..289W}. The relevant Sino-Korean text may be translated as follows: “From the NW to the SW, a red vapour soared and filled the sky.” On this occasion, the red aurora was seen to extend to the south-west of the observing site (Songdo) and hence was seen in both northerly and non-northerly directions. The phase and elongation of the Moon were 0.79 and 125° W, respectively, at the time of this auroral observation \citep[see their Table 2]{2001AnGeo..19..289W}. In this situation, if extensive cloud cover had obscured the northern sky, the aurora would have been seen only in the SW direction. In such a situation, \citetalias{2015AN....336..225N} would have excluded this observation because it would have been a “purely southern sighting”. Indeed, we should note that the MLAT of Chang’an in 776 (37.0° MLAT) and Songdo in 1128 (39.5° MLAT) were not too different, and the “red vapour” extended beyond the zenith of Songdo to the “south-west” in the latter case.

Whether or not aurorae can be seen in non-northerly directions at a particular location clearly depends on how far the auroral display extends equatorward, as already shown above. In this context, it is first necessary to consider if the astronomical event in 776 could not possibly be an aurora. Therefore, we used the computed magnetic latitude of Chang’an in 776, which is 37.0° MLAT. Considering that the equatorward boundaries of the great auroral ovals on 1859 September 1/2 and 1872 February 4/5 came down to 25.1° MLAT and 20.0° MLAT, respectively, it is clearly possible for aurorae to extend southwards beyond the zenith at Chang’an (37.0° MLAT), in the case of extreme events of similar scale.

\section{Whitish Colour of Auroral Displays}
\label{sec:wcad}

This section considers the colour of the astronomical event in 776. The third question posed by \citetalias{doi:10.1002/asna.201512193} was:  “Why was no consideration given to the fact that the white \textit{qi} phenomenon appeared in the eastern to southern direction, not in the north direction?” This question is presumably based on the first “criterion of likeliness”, namely “i) colour”, in \citetalias{2015AN....336..225N}, who assert that some “non-white colours” will satisfy this criterion, while “wordings such as \textit{white}, \textit{brilliant}, or \textit{glow} are neither sufficient nor a contradiction (neutral)”.

\citetalias{2015AN....336..225N} frequently consider a candidate auroral record describing a whitish colour as being “a very weak one without colour” or “white, \textit{i.e.} weak” \citepalias[p.~246]{2015AN....336..225N}. If we interpret literally what \citetalias{2015AN....336..225N} describe here, we do not necessarily need to exclude auroral candidates with whitish colour and their reliability should not necessarily be decreased because of their colour.

Certainly, we have a greater opportunity to see reddish aurorae in low magnetic latitudes. However, when we see an auroral display without enough brightness, it tends to appear whitish to the human eye. This is because of the relative efficiency of rod cells and cone cells in the human eye. In the case of faint aurorae, cone cells responsible for colour do not become active enough and the rod cells responsible for peripheral sight become active and faint aurorae appear “whitish” to the human eye \citep[p.~193]{1993lco..book.....M}. The colour shift of naked-eye observations of the stellar light is also known from \citet[p.~360]{2009ASSL..360.....C} and hence we can expect the same phenomenon as occurs for auroral observations with the naked eye, as discussed by \citet{2017PASJ...69...22T}. In addition, human eyes may not be ``dark adapted'' if the Moon is also in the sky. Therefore, contrast could also contribute to lack of visible colour. In fact, we have several reports of whitish auroral observations in low magnetic latitude areas during the great auroral displays in 1859 \citep[p.~258]{1860AmJS...30..339L}. We also have some simultaneous observations of “whitish” auroras from Spain and Brazil reported by \citet{2017PASJ...69L...1C}. Therefore, regarding whitish aurorae as a “contradiction” can mistakenly remove potential auroral candidates.

Indeed, aurorae are sometimes reported with a “whitish” colour. \citet{2006AdSpR..38..155H} cites a newspaper stating “Strong white-light aurora over southern half [of] the sky and extending to the zenith” at Campbell Town (34°04$'$S, 105°49$'$E) on 1859 August 29. Almost simultaneously, Benjamin V. Marsh at Burlington (40°05$'$N, 74°52$'$W; 51.3° MLAT) reported a curious phenomenon: “Aug. 28th an arch of light rose in the north, passed the zenith and descended to within about 20° of the south horizon by 8h 80m P. M. Soon after this, the whole space overhead was occupied by a dense unbroken cloud of milky whiteness.” \citep[p.~258]{1860AmJS...30..339L}. Marsh, in particular, clearly describes a whitish auroral display down to “about 20° above the south horizon”.

It is well known that a great magnetic disturbance was observed on 1859 August 28/29, prior to the Carrington event on 1/2 September, and that this magnetic disturbance was accompanied by an auroral display \citep{2006AdSpR..38..155H,2006AdSpR..38..130G}. Applying the criteria proposed by \citetalias{2015AN....336..225N}, however, this auroral display does not satisfy either the criterion “i) colour” or the criterion “iii) northern direction”, just like “the white \textit{qi} phenomenon that appeared in the eastern to southern direction” in 776. However,  auroral displays can extend beyond the zenith to the “south”, namely “equatorward” in the northern hemisphere, whereas auroral displays will appear more frequently in the ``south'', namely ``poleward, in the southern hemisphere (see Section~\ref{sec:crcar}). 

\begin{figure}
    \centering
    \includegraphics[width=15.921cm,height=9.4cm]{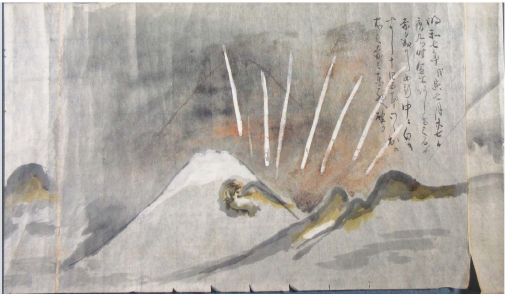}
    \caption{A Japanese drawing of an auroral display observed on 1770 September 16, showing a bright whitish component in the form of multiple streaks of white vapour \citep[courtesy of Shizuoka Prefectural Central Library; see also][]{Hayakawa2017e}.}
    \label{fig:aurora}
\end{figure}

Furthermore, the description ``more than ten streaks (\textit{dao}) of white vapour'' is very similar to the description ``13 streaks of white vapour'', which were observed at \textit{Feicheng} (36°11$'$N, 116°46$'$E; 26.0° MLAT; Qinshigao, v.43, p.~1611) on 1770 September 17. It is well established that a great auroral display was observed throughout East Asia in mid-September, including September 17 \citep{1996QJRAS..37..733W,2017SpWea..15.1373E,Hayakawa2017e}. One night earlier, on 1770 September 16, a Japanese chronicler drew an auroral display above Mt. Fuji, as reproduced in \autoref{fig:aurora} \citep[see][]{Hayakawa2017e}. This figure illustrates the “white vapour” with a rather dim dark-reddish background. During the same night, the \textit{aurora borealis} was observed at several Chinese sites, and the \textit{aurora australis} was observed by Joseph Banks and Sydney Parkinson on board HMS \textit{Endeavour}, captained by James Cook \citep{1996QJRAS..37..733W,Hayakawa2017e}. The reliability of these auroral observations is confirmed by their simultaneous global appearances \citep[\textit{e.g.},][]{2000AnGeo..18....1W}. These examples confirm that some auroral displays are indeed reported to be whitish even in low MLAT during extreme space$-$weather events.

\section{Aurorae Seen Near the Moon}
\label{sec:asnm}

This section address two criticisms that \citetalias{doi:10.1002/asna.201512193} directed at the paper by \citetalias{STEPHENSON20151537}, namely (see \nameref{sec:intro}): “The Chinese record is not consistent with an auroral display, partly because it would be too bright close to the Moon to observe the aurora”, and “The Chinese record could not describe an auroral display because the phenomenon was seen above the Moon”. Both of these criticisms might have been developed from criterion iv) in the paper by \citetalias{2015AN....336..225N}.

However, whether or not an aurora is overshined by the Moon depends on the lunar phase, the brightness of the aurora, and the angular distance between the Moon and the sky positions occupied by auroral emission. \citet{2017SpWea..15.1373E} have examined reports of extremely bright aurorae in low magnetic latitudes ($\approx$ 25° MLAT) on 1770 September 17, and concluded that these auroral displays should be classified as Class IV of the International Brightness Coefficient (IBC) in terms of their brightness -- equivalent to that of the full Moon per unit area on the ground \citep[pp.~124--125]{Chamberlain1961} -- under realistic extreme fluxes of high-intensity, low-energy electrons.

In fact, there are some compelling examples of auroral displays observed during the night at the time of full Moon. \citet{1910ApJ....31..208B} reports an auroral observation at full Moon, observed at the Yerkes Observatory (42°34$'$N, 88°33$'$W) on ``September 4 (1909). 7h 30m: Bright aurora with nearly full moon. ... 10h40m: The arch was very bright in spite of a bright moon.'' The contemporary magnetic latitude of Yerkes Observatory is computed to be 53.1° MLAT, according to the IGRF model.

On 1847 October 24, an auroral display was observed during the night with full Moon (lunar phase = 0.986). One of the auroral observations was made at Ramsgate (51°20$'$N, 1°25$'$E; 54.6° MLAT) and described as follows: ``It streamed out of the sky from N.W. to N.E. towards the zenith, flickering like pale gaslight at its source, and gradually deepening as it rushed into the firmament, till it assumed so deep a crimson at times over head that our pier light which is a brilliant red with catoptric lenses and reflectors, paled into an orange colour beneath it; the moon's brilliancy had no effect in dimming the lustre of the Aurora, it advanced and retired within a few degrees of her disc, while every star and constellation above it, was seen as through a thin gauze'' \citep[p.~643]{Martin1847}. This observation shows explicitly that the aurora came to a few degrees above the Moon’s disk. During the time of this observation at Ramsgate, \citet{doi:10.1080/14786444708645871} also observed the aurora but ``did not observe any halo around the Moon at any time'' at Blackheath (51°28$'$N, 0°00$'$E; 55.0° MLAT). The magnetometer at Greenwich on the same night shows an intensive magnetic disturbance ($\Delta H \approx -1100$ nT), as shown in Figure 8 of \citet{2013JSWSC...3A..31C}, which supports the existence of the brilliant aurora. This is indeed a good example, suggesting that an auroral display can be seen around the time of the full Moon if it is bright enough.

This is not the only example of an auroral display during the night at the time of full Moon. Robert \citet[pp.~9--11]{Snow1842} reported eyewitness auroral observations during the night at full Moon at Ashurst (51°16$'$N, 0°01$'$W; 55.0° MLAT) on 1837 February 18 and at Dulwich Wood, London (51°26$'$N, 0°04$'$W; 55.1° MLAT) on 1837 November 12. The lunar phases were 0.974 and 0.994 respectively. \citet[p.~178]{Olmsted1837} reported another eyewitness auroral observation during the night with almost full Moon (lunar phase 0.834) at Yale (41°19$'$N, 72°55$'$W; 52.4° MLAT) on 1837 January 25.

During the Carrington Event, as well, the aurora was reported as being “bright as full moon” \citep{2006AdSpR..38..145G}. Reports in American newspapers have frequently compared the brightness of auroral displays with that of the full Moon \citep{2017PASJ...69...22T}. In the case of the auroral display on 1859 August 28/29, the Washington Daily National Intelligencer (1859-08-31) reported ``The light appeared in streams, sometimes of a pure milky whiteness and sometimes of a light crimson. The white and rose-red waves of light as they swept to and from the corona were beautiful beyond description, and a friend nearby us, while looking to the zenith with the whole heavens and earth lighted up at a greater brilliancy than is afforded by the full moon, said that it was like resting beneath the wings of the Almighty''. In the case of the auroral display on 1859 September 1/2, the Alabama Educational Journal (v.1, p.~370) reported ``Sheets of the same white luminous cloud again illuminated the sky, producing about the same amount of light as the full moon, and the night became almost as the day''. In addition, a report in the Baltimore American and Commercial Advertiser describes the auroral observation on 1859 September 1/2 as ``The light was greater than that of the moon at its full, but had an indescribable softness and delicacy''. These reports seem to emphasise that the extremely bright aurora was observed not only during the Carrington event on 1859 September 1/2, but also during the preceding magnetic storm on 1859 August 28/29.

The reports presented in this section, together with those in Section~\ref{sec:crlhh}, show that auroral displays can be seen above the Moon if they are bright enough. The lunar phase was 0.99 on 776 January 12 and hence the Moon was almost full.

\section{The Lunar Halo Hypothesis}
\label{sec:lhh}

This section addresses the second question posed by \citetalias{doi:10.1002/asna.201512193}; namely, ``Why was the lunar halo hypothesis not considered by \citetalias{STEPHENSON20151537}?'' The first point to be emphasised is that the description of the phenomenon observed on the night of AD 776 January 12/13, in terms of more than ten steaks of white vapour, closely resembles the descriptions given in other records that have been classified as auroral observations in several catalogues. \citetalias{2015AN....336..225N} (p.~240) claimed ``The \textit{moon} is otherwise never mentioned close together with true aurorae in Chinese reports, \textit{silk} is only very rarely mentioned''. However, their claim is hardly supported by the Chinese historical documents themselves. Even in \citet{1995caof.book.....Y}, some auroral candidates were explicitly mentioned in association with the Moon, for example, on 1417 December 31, 1634 March 16, or 1643 September 26. They are described as ``a glorious blaze of fire'', ``a red vapour'', and ``a golden light'', respectively. Likewise, ``silk'' is mentioned up to 15 times. One of these reports described ``a white vapour like silk'' on 1101 January 31, confirmed by simultaneous observations in East Asia \citep[\textit{Liaoshi}, v.26, p.~314;][]{2000AnGeo..18....1W}. Another report tells us ``more than 10 bands of white vapour'' penetrated a red cloud ``like unspun silk'' on 1130 June 18 (\textit{Songshi}, v.60, p.~1314). Therefore, it was natural and appropriate to consider an auroral interpretation first.

Moreover, there are strong scientific reasons for questioning the lunar halo hypothesis advanced by \citetalias{2015AN....336..225N} and \citetalias{doi:10.1002/asna.201512193}. It is crucially important to estimate the spatial extent of the ``more than ten streaks of white vapour'' seen on the night of 776 January 12/13 in terms of the asterisms that they penetrated, as observed from Chang’an (Section~\ref{sec:cht}), without making any prior assumption that these streaks of white vapour represent an auroral display. If the ``more than ten streaks of white vapour'' penetrated the eight Chinese asterisms (star groups) simultaneously, soon after moonrise, then the spatial extent of the white vapour would have been much greater than the spatial extent of any simple lunar halo.

\begin{figure}
    \centering
    \subfloat{%
        \label{fig:chart_moonrise}%
        \includegraphics[width=0.75\textwidth]{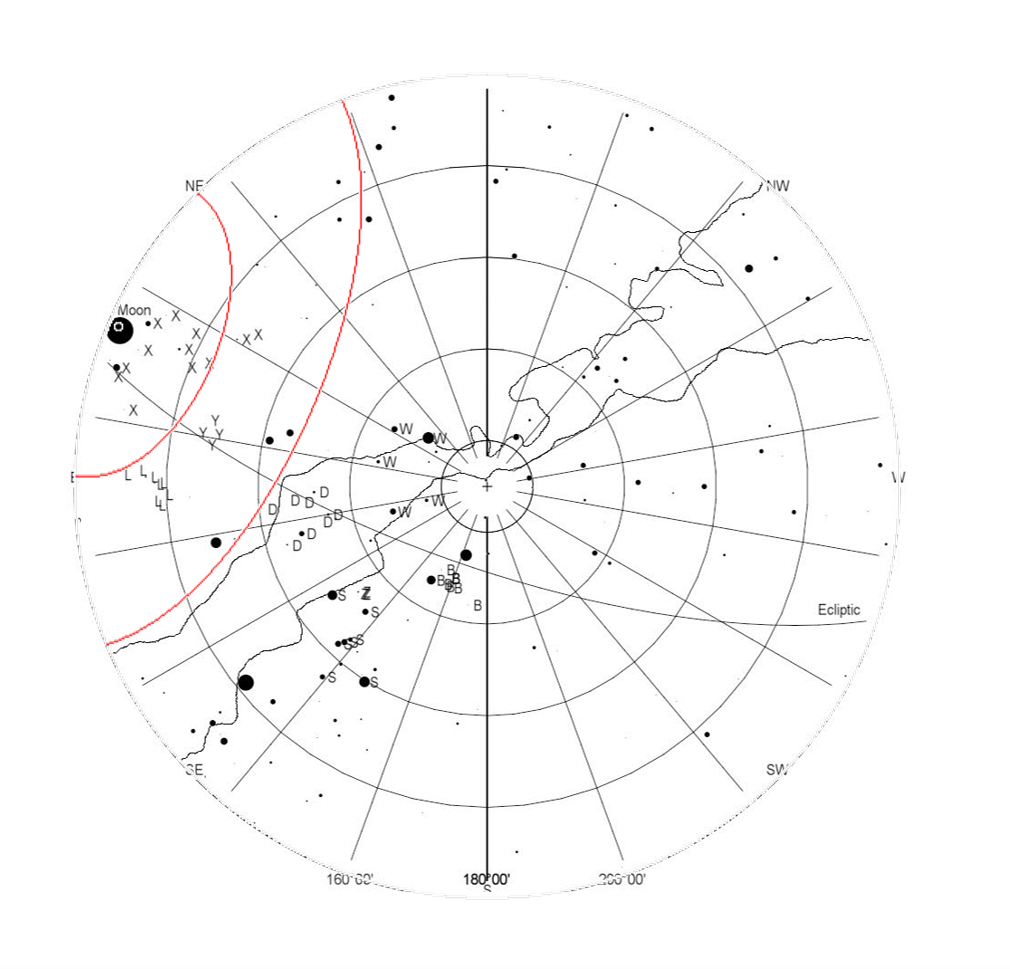}}
    \qquad
    \subfloat{%
        \label{fig:chart_thirdwatch}%
        \includegraphics[width=0.75\textwidth]{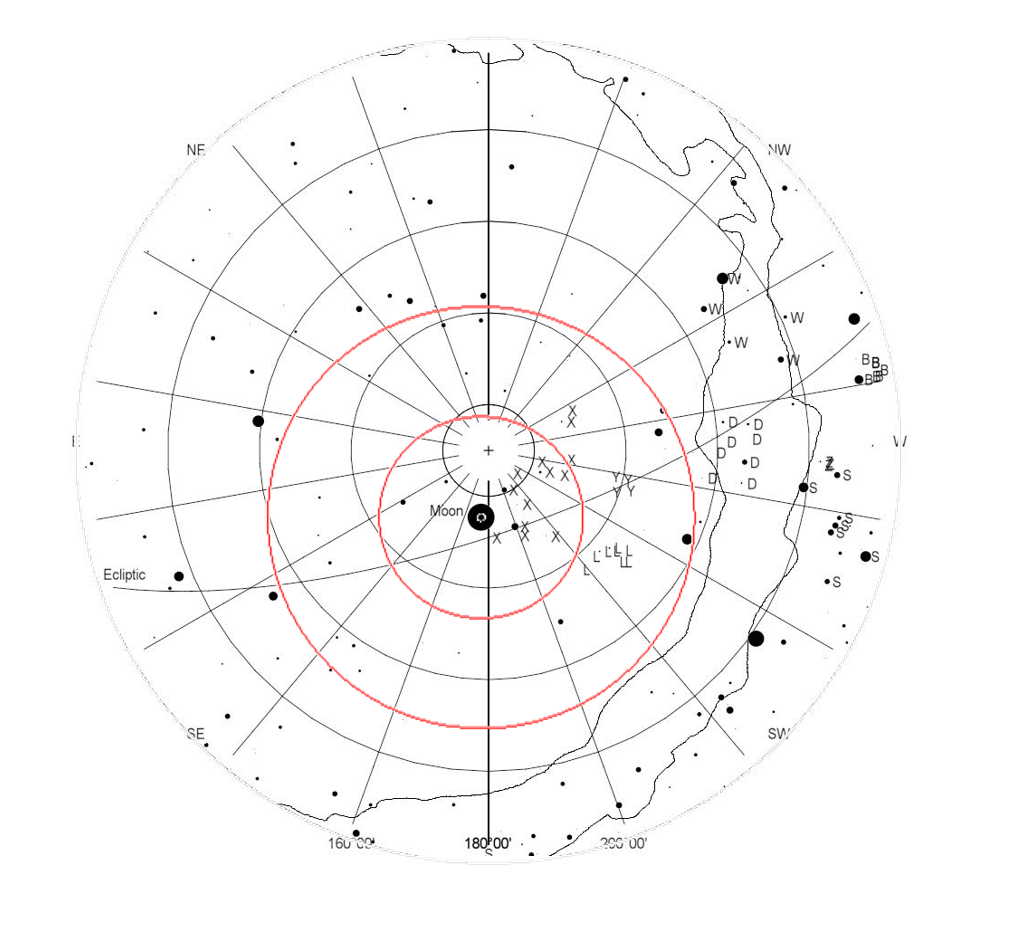}}
    \caption{(a, top) Star chart showing the positions of lunar halos with radii of 22° and 46° at moonrise (18.72 h LT on AD 776 January 12), as viewed from Chang’an. (b, bottom) Similar star chart showing the positions of the lunar halos and the 56 representative stars at the end of the third watch (1.50 h LT on AD 776 January 13). All-sky view (altitude-azimuth coordinate system) with geographic north at the top and zenith at the centre. The positions of the 56 stars are indicated by the initial letter (in parentheses) of the Chinese asterism in which they are located. (\textit{Wuche} (W), \textit{Dongjing} (D), \textit{Yugui} (Y), \textit{Zui}[\textit{xi}] (Z), \textit{Shen} (S), \textit{Bi} (B), \textit{Liu} (L), and \textit{Xuanyuan} (X)).}
    \label{fig:charts}
\end{figure}

The historical records in the \textit{Jiu Tangshu} and the \textit{Xin Tangshu} note that, at night, in the eastern direction above the Moon there were more than ten streaks of white vapour like unspun silk. They penetrated [the star groups] \textit{Wuche} (W), \textit{Dongjing} (D), \textit{Yugui} (Y), \textit{Zui}[\textit{xi}] (Z), \textit{Shen} (S), \textit{Bi} (B), \textit{Liu} (L), and \textit{Xuanyuan} (X) (see Section~\ref{sec:cht}). Figure~\ref{fig:chart_moonrise} indicates the spatial extent of the eight Chinese asterisms on AD January 12, at a Local Time (LT) of 18.72 h (18:43), as viewed from Chang’an (34°14$'$N, 108°56$'$E or, equivalently, 34.23°N, 108.93°E). The time of moonrise (18.72 h LT) has deliberately been selected to give a lunar altitude of +2.2°, to allow for possible obscuration of the Moon by hills near the horizon. The time of moonrise for an idealised smooth horizon is 18.68 h LT, which corresponds to a Terrestrial Time of 12.50 h. The value of Delta T (DT), the difference between Terrestrial Time (TT) and Universal Time (UT), is taken to be 3000 seconds for AD 776 January \citep{2016RSPSA.47260404S}. At moonrise (18.72 h LT), the azimuth of the Moon (measured E of N) was 66.6°. The Moon was then 158°W of the Sun. Figure~\ref{fig:chart_moonrise} also shows the altitudes and azimuths for 56 representative stars within the eight Chinese asterisms (star groups) at 18.72 h LT. The location of the 56 stars are identified by the initial letter of the Chinese asterism (W, D, Y, Z, S, B, L, and X) within which they are located. Also shown in Figure~\ref{fig:chart_moonrise} are the positions of lunar halos with radii of 22° and 46° at 18.72 h LT.

It is clear from Figure~\ref{fig:chart_moonrise} that a lunar halo with a radius of 22° would have penetrated the Chinese asterism \textit{Xuanyuan} (X) on AD 776 January 12 at 18.72 h LT. Similarly, it is perhaps conceivable that a lunar halo with a radius of 46° could have been perceived to have penetrated the Chinese asterisms \textit{Yuqui} (Y) and \textit{Liu} (L), and just possibly \textit{Dongjing} (D) as well. However, it is clearly impossible for a lunar halo with a radius of either 22° or 46° to have penetrated the Chinese asterisms \textit{Wuche} (W), \textit{Zui}[\textit{xi}] (Z), \textit{Shen }(S), and \textit{Bi} (B). Moreover, it should be noted that the streaks of white vapour are completely free from the influence of solar light as the altitude of the Sun was $-$19.8° at moonrise, while at astronomical twilight the altitude of the Sun was less than $-$18.0°. Therefore, an explanation of the more than ten streaks of white vapour observed at Chang’an at 18.72 h LT on AD 776 January 12 in terms of a simple lunar halo is untenable.

The historical record in the \textit{Jiu Tangshu} notes that the more than ten streaks of white vapour, like unspun silk, disappeared after the third watch (about 1:30 LT). On AD 776 January 12 at 19.5 h TT (January 13 at 1.50 h LT or 1:30 LT), at Chang’an (34° 14$'$N, 108°56$'$E or, equivalently, 34.23°N, 108.93°E), the lunar azimuth (measured E of N) was 173.8° and the altitude was + 75.4°. The difference between Terrestrial Time (TT) and Universal Time (UT) is again taken to be 3000 seconds. Figure~\ref{fig:chart_thirdwatch} shows the positions of lunar halos with radii of 22° and 46°, and the altitudes and azimuths of the 56 representative stars within the eight Chinese asterisms, at 1.50 h LT on AD 776 January 13.

It is clear from Figure~\ref{fig:chart_thirdwatch} that a lunar halo with a radius of 22° would have penetrated the Chinese asterism \textit{Xuanyuan} (X) on AD 776 January 13 at 1.50 h LT. Once again, it is perhaps conceivable that a lunar halo with a radius of 46° could have been perceived to have penetrated the Chinese asterisms \textit{Yuqui} (Y) and \textit{Liu} (L), and just possibly \textit{Dongjing} (D) as well. However, it is clearly impossible for a lunar halo with a radius of either 22° or 46° to have penetrated the Chinese asterisms \textit{Wuche} (W), \textit{Zuixi} (Z), \textit{Shen }(S), and \textit{Bi} (B) at 1.50 h LT, assuming that the streaks of white vapour persisted until the end of the third watch. Therefore, even if the more than ten streaks of white vapour observed at Chang’an persisted until 1.50 h LT on AD 776 January 13, an explanation in terms of a simple lunar halo is again untenable.

As shown in \autoref{fig:charts}, the wide distributions of the eight Chinese asterisms penetrated by the ``more than ten streaks of white vapour'' clearly contradict the assertions by \citetalias{2015AN....336..225N} and \citetalias{doi:10.1002/asna.201512193}, namely that the white vapour appeared ``\textit{only} above the moon'' \citepalias[p.~240]{2015AN....336..225N} and that it was ``too bright close to the moon'' \citepalias[p.~535]{doi:10.1002/asna.201512193}. There is no indication in these figures that the white vapour was concentrated around the Moon. Indeed, the precise descriptions of the ``more than ten streaks of white vapour'' in the original records indicate that they were never ``\textit{only} above the moon'' but extended to regions of the night sky well outside and beyond lunar halos having radii of 22° or 46° \citep[\textit{e.g.},][]{1993lco..book.....M}.

Additionally, the streaks of white vapour lasted more than 7 hours, which is slightly longer than the duration expected of lunar halo displays that ``last tens of minutes and even hours'' \citep[p.~381]{1993lco..book.....M}. Often auroral displays last from sunset to sunrise, particularly when multiple magnetic storms occur, as in the case of the extreme space-weather events in 1859 \citep[\textit{e.g.},][]{2006AdSpR..38..130G}.

Moreover, while solar halos sometimes have a radius greater than 46° \citep[\textit{e.g.},][pp.~208--232]{1993lco..book.....M}, this is not normally the case for lunar halos because of their lesser brightness. \citet[p.~209]{1993lco..book.....M} states that most of the larger halo effects are caused by the Sun and ``those belonging to the moon are much fainter and their colours are practically imperceptible'' \citep[p.~209]{1993lco..book.....M}. Furthermore, the white vapour did not ``circle'' or ``penetrate'' the Moon but was ``above'' the Moon: it comprised ``more than ten streaks''. In this case, it is unlikely that the Chinese record describes a more extensive lunar halo (like a parhelic arc) with a radius much greater than 46°. Therefore, the large spatial extension of the white vapour is not considered to be a large halo associated with the Moon.

\begin{figure}
    \centering
    \subfloat{\includegraphics[width=0.47\textwidth,height=13cm]{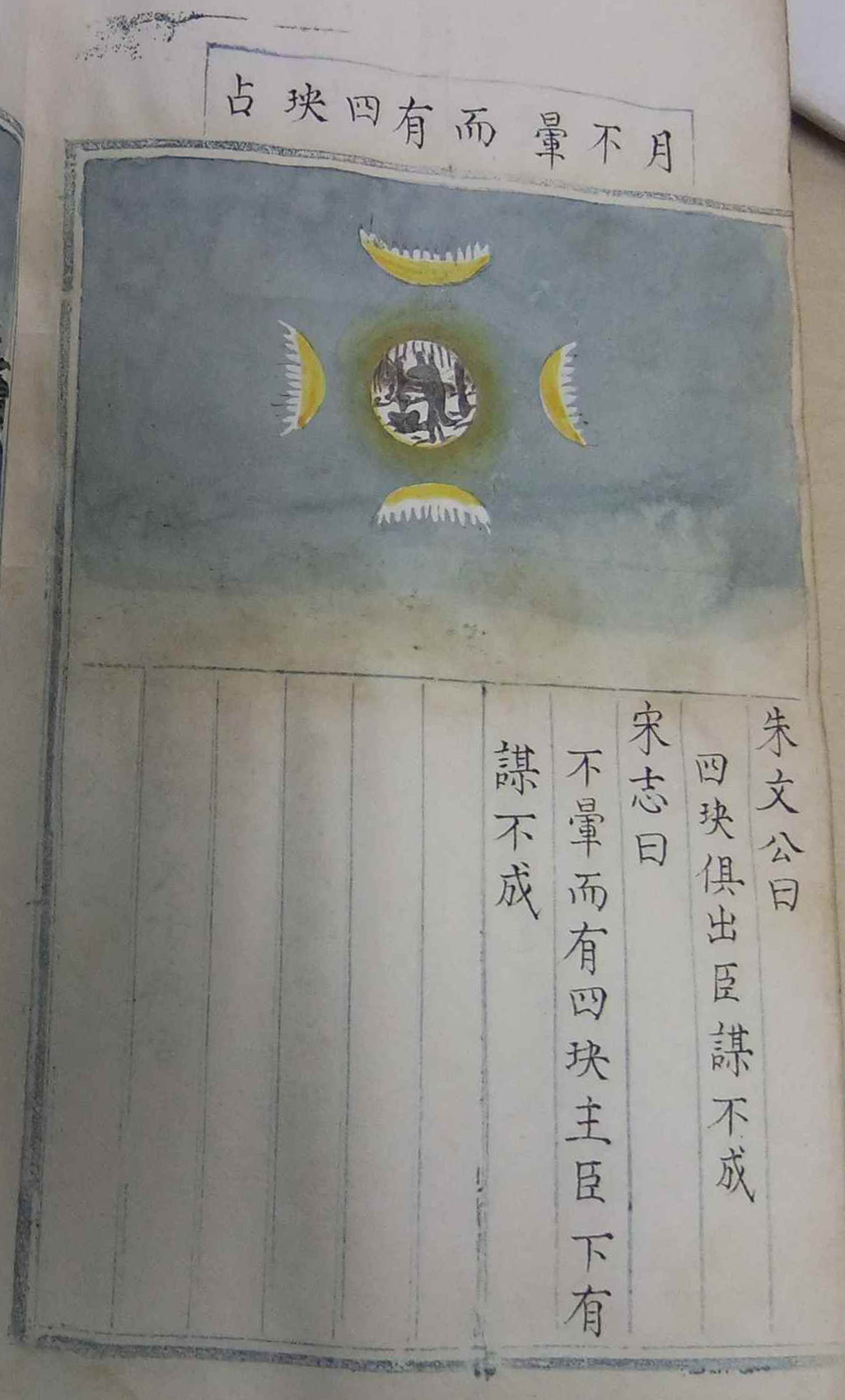}}
    \subfloat{\includegraphics[width=0.47\textwidth,height=13cm]{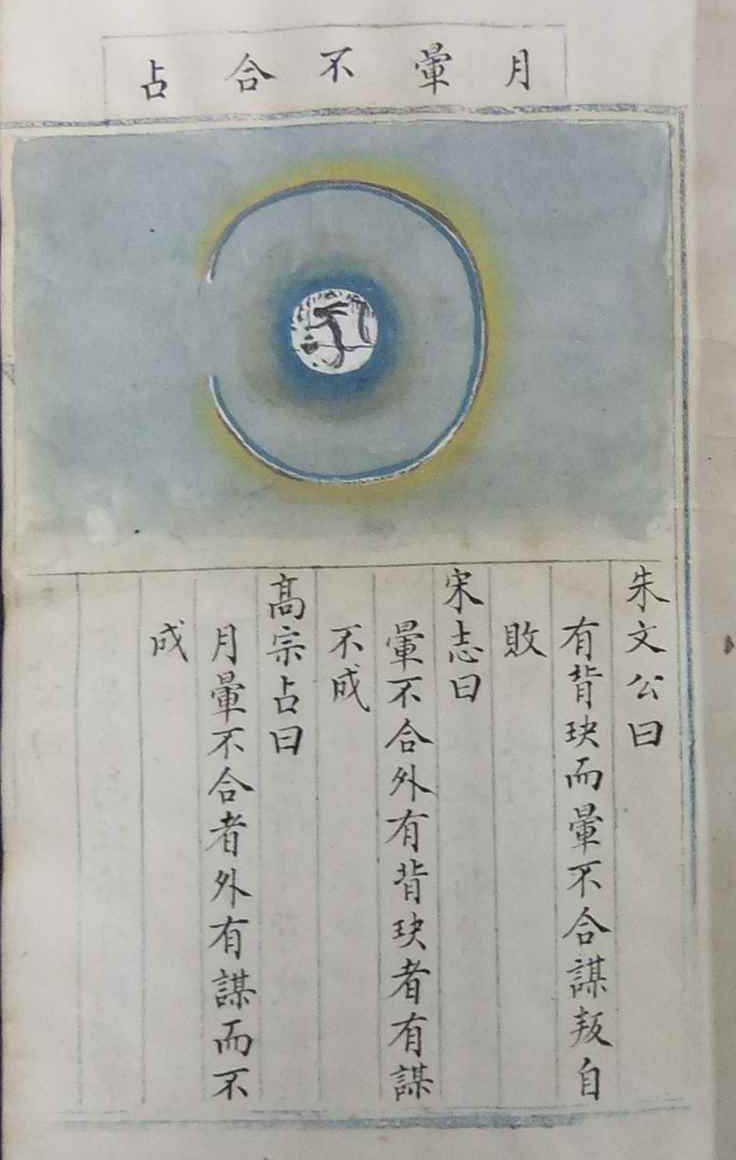}}
    \caption{Lunar halos in Chinese astro-omenological text (\Chi{天元玉暦祥異賦};\textit{Tianyuan Yulì Xiangyìfu}, v.3, f.42b and v.4, f.8a; MS 2573, The National Archives of Japan).}
    \label{fig:halo}
\end{figure}

It also seems highly significant that the Chinese astronomers themselves did not use any technical terms for a lunar halo to describe the astronomical event on AD 776 January 12/13. \citet{1959Wthr...14..124P} have noted that the section on the ``Ten Haloes'' (\textit{shi yun}) in one of the astronomical chapters of the \textit{Chin Shu} (Official History of the Chin Dynasty), completed about AD 635, contains some twenty-six technical terms that can be identified with almost every component of the solar halo system. It seems likely that the same nomenclature would also have been used to describe lunar haloes, as in the \textit{Tianyuan Yulì Xiangyìfu} --- a Chinese astrono-omenological manual used by Chinese court astronomers \citep{sasaki2013}, as reproduced in \autoref{fig:halo}. Indeed, as mentioned in Section~\ref{sec:cht}, it is inferred from the records that the observer had very detailed knowledge and experience of astronomical observations, as the descriptions were remarkably detailed and have even captured constellations with only faint stars like \textit{Yugui}(${\geq}$ 3.9), \textit{Zuixi}(${\geq}$ 3.4), or \textit{Liu}(${\geq}$ 3.1). With such an array of technical terms available since about AD 635, to describe solar (and hence lunar) halos, it is unlikely that the Chinese astronomers would have overlooked any simple lunar halo, or even a relatively complex one, on the night of AD 776 January 12/13.

\section{Justification for an Auroral Interpretation of the Chinese Record}
\label{sec:jaicr}

As discussed in the preceding section, it is extremely difficult, if not impossible, to explain the Chinese record of AD 776 January 12/13 in terms of a simple, or even complex, lunar halo. Conversely, an auroral interpretation of the Chinese observation on AD 776 January 12/13 at 18.72 h LT can be justified as follows.

Using the archaeomagnetic model derived by \citet{Korte2011}, we have estimated the dipole declination and inclination at Chang’an (34°14$'$N, 108°56$'$E) in AD 776 to be $-$2.24° and 55.9°, respectively. 

\begin{figure}
    \centering
    \subfloat{%
        \label{fig:rays_moonrise}%
        \includegraphics[width=0.75\textwidth]{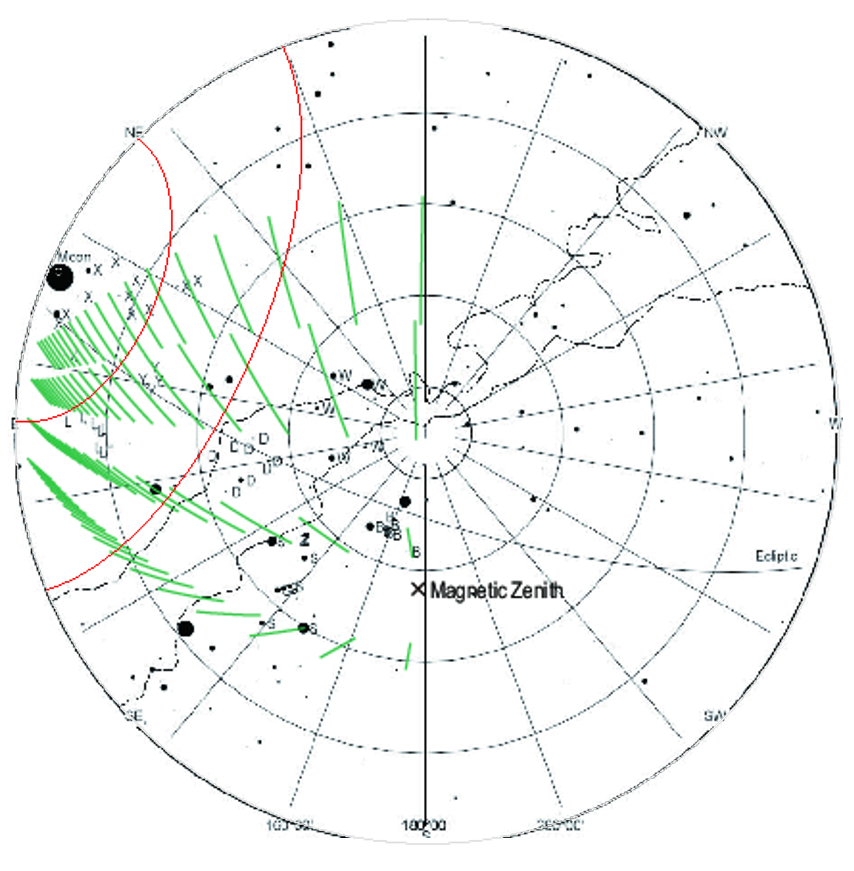}}
    \qquad
    \subfloat{%
        \label{fig:rays_thirdwatch}
        \includegraphics[width=0.75\textwidth]{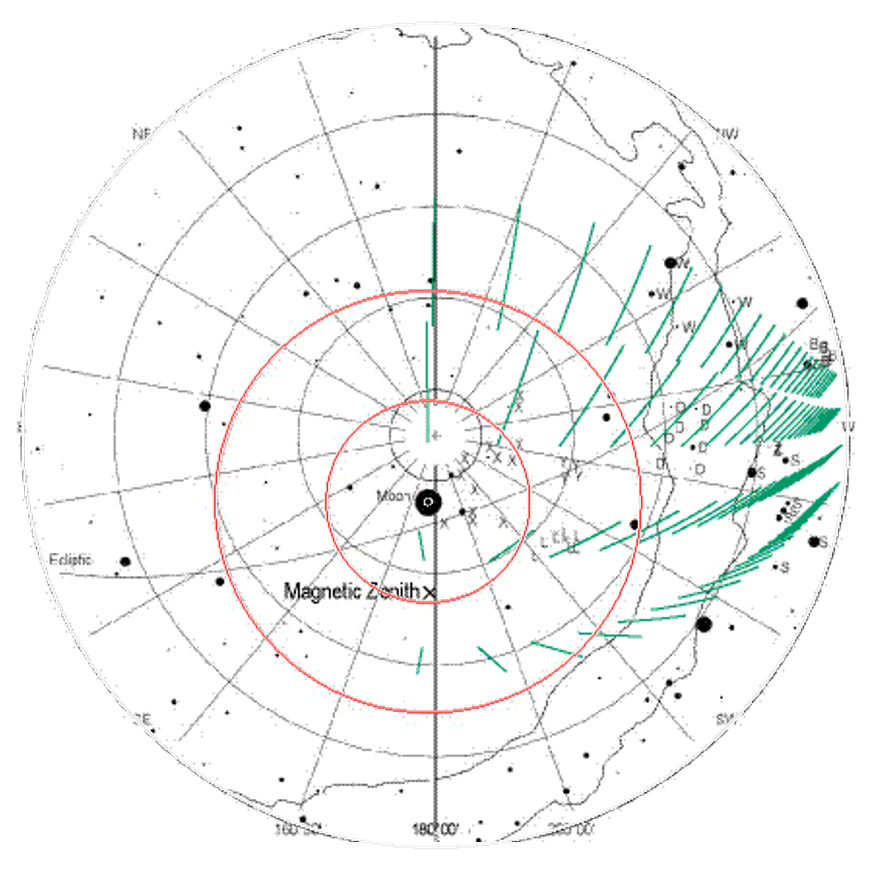}}
    \caption{(a, top) All-sky views of the auroral ray structures over Chang’an: after  moonrise (18.72 h LT on 12 January 776). (b, bottom) All-sky views of the auroral ray structures over Chang’an: after the third watch (1.50 h LT on 13 January 776). The ray structures are assumed to extend from 97 km to 170 km altitudes along the dipolar magnetic field lines. The geographic north is to the top. The $\times$ symbol indicates the local magnetic north estimated by the CALS3k.4b model.}
    \label{fig:rays}
\end{figure}

When precipitating electrons excite an atomic oxygen to the O (\textsuperscript{1}S) state, the transition OI (\textsuperscript{1}D-\textsuperscript{1}S) results in emission at 557.7 nm \citep{1975RvGSP..13..201R}. The excitation energy of the OI (\textsuperscript{1}S) is 4.19 eV. \citet{2017SpWea..15.1373E} calculated volume emission rates at 557.7 nm and 630.0 nm for the precipitation electron flux observed at mid-latitudes during the intense storm on 1989 March 13/14 (with a minimum Dst of $-$589 nT). According to their calculation, the highest volume emission rate takes place at 109 km altitude. One-tenth of the highest volume emission rate takes place at 97 km altitude and 170 km altitude. We assume that an observer in Chang’an recorded ray structures extending from 97 km to 170 km along a magnetic field line. Figure~\ref{fig:rays_moonrise} shows an all-sky view of the field-aligned ray structures over Chang’an at 18.72 h LT on AD 776 January 12. The centre of the view corresponds to the zenith at Chang’an. We arbitrarily chose the magnetic field lines at \textit{L}=1.55, 1.58, 1.61, and 1.64 (corresponding to 36.6°, 37.3°, 38.0°, and 38.7° invariant latitudes) at magnetic longitudes from 31° to 40°. The magnetic field lines are assumed to be dipolar. It is clearly seen that the ray structures traverse the eight Chinese star groups above the Moon. To satisfy this condition, the ray structures must appear on the magnetic field lines between \textit{L}=1.55 and 1.64 (corresponding to 36.6° and 38.7° ILATs). Figure~\ref{fig:rays_thirdwatch} is the same as figure~\ref{fig:rays_moonrise} with the time adjusted to 1.5h LT on AD 776 January 13. The magnetic field lines at magnetic longitudes from 21° to 31° are drawn. Again, the ray structures traverse the eight Chinese star groups.

With the preceding assumptions, the equatorward boundary of the auroral oval is located at invariant latitude (ILAT) of 36.6°, or less, and the poleward boundary is located at an invariant latitude (ILAT) of 38.7°, or more. Therefore, the equatorward boundary of the auroral oval on AD 776 January 12/13 is comparable with the equatorward boundary of the auroral oval during known extreme magnetic storms, such as the extreme events in 1859 (30.8° ILAT or 28.5° ILAT) and in 1872 (24.2° ILAT) \citep{2018ApJ...862...15H,2018arXiv181102786H}. Therefore, an auroral interpretation of the Chinese record of AD 776 January 12/13 appears to be entirely feasible, according to the corresponding records for known extreme space--weather events.

\section{Conclusions}
\label{sec:conc}

In this article, we have reinvestigated the Chinese astronomical record (auroral candidate) on AD 776 January 12/13 and considered if its interpretation as an auroral display \citep{STEPHENSON20151537} can be replaced by the ``lunar halo hypothesis'' \citepalias{2015AN....336..225N,doi:10.1002/asna.201512193}, according to the recently ``established'' auroral criteria propounded by \citetalias{2015AN....336..225N}. Our research has revealed several counter-examples, including the auroral beam on 1882 November 17 \citep[\autoref{fig:capron}]{Capron1883}, which are not in agreement with the auroral ``criteria'' in \citetalias{2015AN....336..225N}.

Their criterion ``i) colour'' fails to consider the detection threshold of auroral colour in the human eye, and cannot explain the whitish auroral displays observed during extreme space--weather events, like the one on 1859 August 28/29, prior to the Carrington Event. Their criterion ``iii) northern direction'' fails to take into account not only the aurora australis in the southern hemisphere but also auroral displays extending down to low magnetic latitudes during extreme space--weather events. Moreover, Chang’an was much closer to the magnetic north pole in 776 (37.0° MLAT) than at modern times \citep[\textit{e.g.}, 24.5° MLAT in 2014; see,][]{2015EP&S...67...79T}.

\citetalias{2015AN....336..225N} and \citetalias{doi:10.1002/asna.201512193} have proposed the ``lunar halo hypothesis'' for the AD 776 event, on the basis that the white vapour appeared ``\textit{only} above the moon'' \citepalias[p.~240]{2015AN....336..225N} and that it was ``too bright close to the moon'' \citepalias[p.~535]{doi:10.1002/asna.201512193}. These restrictions might possibly be related to the criterion ``iv) darkness'', although this conjecture is not clarified explicitly in \citetalias{2015AN....336..225N}. However, the named Chinese asterisms (constellations) penetrated by the ``white vapour'', according to the original records, indicate that this vapour was not seen ``\textit{only} above the moon'' but extended to regions of the night sky well beyond and outside lunar halos with radii of 22° or 46° \citep[\textit{e.g.},][]{1993lco..book.....M}. Moreover, the Chinese astronomers had already introduced some 26 technical terms for solar halos (and hence lunar halos too) by AD 635 \citep{1959Wthr...14..124P} but they did not use such technical terms to describe the astronomical event in AD 776. We have also confirmed that auroral displays can be visible even near the full Moon if they are sufficiently bright, as on 1847 October 24.

In short, we have confirmed that the recent auroral ``criteria'' published by \citetalias{2015AN....336..225N}, and the resultant lunar halo hypothesis propounded in \citetalias{2015AN....336..225N} and \citetalias{doi:10.1002/asna.201512193}, are not based on solid scientific foundations. Furthermore, these authors’ discussions on the level of solar activity around the time of the \textsuperscript{14}C event in AD 775 almost retraces the arguments advanced by \citetalias{STEPHENSON20151537} and their additional conclusions are incorrect, now we have indicated that their auroral ``criteria'' are subject to numerous counter-examples.

While there are firm examples of halo observations initially being misinterpreted as auroral displays, such discussions must be conducted carefully in comparison with modern scientific knowledge, as done by \citet{2017SoPh..292...15U}.

The white vapour penetrating the eight Chinese asterisms can be reproduced if we consider the auroral-ray structures at altitudes between 97 km and 170 km, along the magnetic field lines between \textit{L}=1.55 and 1.64 (corresponding to 36.6° and 38.7° ILATs) as shown in \autoref{fig:rays}. Most of the area penetrated by the ``white vapour'' does not even extend beyond the magnetic zenith and this is well within the spatial extent of known extreme magnetic storms, such as those on 1859 September 1/2 (30.8° or 28.5° ILAT) and 1872 February 4 (24.2° ILAT) in early modern observations.

It is remarkable that an observation made almost one-and-a-quarter millennia ago, which is recorded in Chinese history, can be subjected to so much careful scientific analysis and can be compared meaningfully with much more modern and detailed observations.

The event discussed in this article  took place approximately 1.5 years after the boreal summer of AD 774, the expected onset of the extreme solar energetic particle (SEP) event  \citep{2017JASTP.152...50S,Buntgen2018,Uusitalo2018}. Therefore, the auroral display on the night of AD 776 January 12/13 is not directly associated with the extreme SEP event \citep[see also][]{STEPHENSON20151537} but indicates an enhanced level of solar activity around AD 774/775.

We need further knowledge and understanding of the level of auroral activity during extreme space--weather events, in order to evaluate the potential value of candidate auroral records in historical sources, which let us trace the history of visible aurorae back to the 6th century BC \citep{2004A&G....45f..15S,2006AdSpR..38..136S,2016PASJ...68...99H,2017PASJ...69...17H}. These candidate auroral reports in historical sources are of great importance because systematic magnetic observations only began in the mid-19th century, and extreme space--weather events pose a critical threat to modern civilization in the Space Age \citep{Hapgood2011,NAP12507}.

\section*{Acknowledgements}

This article is supported by the UK Solar System Data Centre; Mission Research Projects of RISH (Kyoto University); and grants-in-aid JP17J06954, JP15H05815, JP15H03732, and JP15H05816. We thank the Shizuoka Prefectural Central Library and the National Archives of Japan for providing permission to research their historical manuscripts, and S. Sasaki, T. Takeda, and S. Hayakawa for their advice on the astronomical knowledge in historical Chinese astronomy and the calculation procedure for the reconstruction of auroral ovals.

\section*{Disclosure of Potential Conflicts of Interest.}
The authors declare that they have no conflicts of interest.

\appendix
\section{References for Historical Documents}

\textit{Alabama Educational Journal: A Magazine of Education}, v.1, p.~370, 1859.
\\
\textit{Baltimore American and Commercial Advertiser}, September 3, 1959.
\\
\textit{Fujiyama Houei Funka-no-Zu: 5}, MS 03036, Kinsei Shashinshu 08-K314, Center for Information of History and Culture, Shizuoka Prefectural Central Library.
\\
\textit{Goryeosa}: Jeong In-ji (ed.), Goryeosa, Seoul: Asea Muhwasa, 1990 (in Sino-Korean).
\\
\textit{Jiutangshu}: Liu Xu (ed.) Jiutangshu, Beijing, 1975 (in Chinese).
\\
\textit{Liaoshi}:Tuoto \textit{et al.} (eds), Liaoshi, Beijing, 1995.
\\
\textit{Qinshigao}:  Zhao Erxun (ed.), Qinshigao, Beijing, 1976
\\
\textit{Tianyuan Yulì Xiangyìfu}, MS 2573, The National Archives of Japan (in Chinese).
\\
\textit{Washington Daily National Intelligencer}, September 6, 1859.
\\
\textit{Xintangshu}: Ouyang Xiu and Song Qi (eds.), Xintangshu, Beijing, 1975 (in Chinese).
\\

\bibliographystyle{spr-mp-sola}

\end{article}
\end{document}